\documentclass[12pt]{article}%
\usepackage{amsmath}
\usepackage{amsfonts}
\usepackage{amssymb}
\usepackage{graphicx}%
\providecommand{\U}[1]{\protect\rule{.1in}{.1in}}

\begin{document}

\title{Electron-Phonon Mass Enhancement in Multi-Layers}
\author{Gerd Bergmann\\Department of Physics\\University of Southern California\\Los Angeles, California 90089-0484\\e-mail: bergmann@usc.edu}
\date{\today}
\maketitle

\begin{abstract}
A strong electron-phonon interaction in a metal increases the electron density
of states in the vicinity of the Fermi energy dramatically. This phenomenon is
called electron-phonon mass enhancement. In this paper the question is
investigated whether the mass enhancement can be manipulated in multi-layers
of two metals with strong and weak electron-phonon interaction. A rich
behavior is observed for different thickness ranges of the layers. For thin
layers one observes a rather homogeneous averaged enhancement. However, for an
intermediate thickness range the mass enhancement is highly anisotropic, i.e.
direction dependent, as well as position dependent. For large layer
thicknesses one obtains the bulk behavior for each metal.


\end{abstract}

\section{Introduction}

A strong electron-phonon interaction (EPI) alters the electronic properties of
a metal rather dramatically \cite{S68}, \cite{G42}. It enhances the electron
density of states in the vicinity of the Fermi energy. The overall strength of
the EPI is summed up in a parameter $\lambda$. Lead is a good example for a
metal with strong EPI having a $\lambda$ value of $\lambda\thickapprox1.6$. As
a consequence the electron density of states at the Fermi energy is enhanced
by the factor $Z=\left(  1+\lambda\right)  \thickapprox2.6$. In addition the
Fermi velocity $v_{F}^{\ast}=\hbar k_{F}/m^{\ast}$ is reduced by the factor
$Z$ corresponding to an enhancement $Z$ of the mass. (Therefore this effect is
often called "mass enhancement"). Since metals with strong EPI are generally
superconducting a number of superconducting properties are also enhanced, for
example the upper critical field $B_{c2}$.

Although there have been many experiments which investigated the properties of
double and multi-layers of a metal S with strong EPI and a metal W with weak
EPI, the author is not aware of any theoretical investigation of how the
contact between S and W influences the mass enhancement in S or W. Such an
investigation is the goal of this paper.

In a metal with a strong electron-phonon interaction the Fermi surface is not
sharp even at zero temperature. This is shown in Fig.1. The occupation below
the Fermi momentum (we discuss here the case of clean metals) is less than one
and above the Fermi energy the occupation is not zero but finite. At the Fermi
energy the occupation drops by the value $Z^{-1}$. The reason for this
distribution is the following. An electron $\mathbf{k}^{\prime}$ below the
Fermi energy can virtually emit a phonon $\left(  \mathbf{q},\lambda\right)  $
and make a transition into a satellite state $\mathbf{k}^{\prime\prime}$ above
the Fermi energy. This process does not fulfill energy conservation since the
satellite state $c_{\mathbf{k}^{\prime\prime}}^{\ast}a_{\mathbf{q},\lambda
}^{\ast}$ has an excess energy of $\Delta E=\varepsilon_{k^{\prime\prime}%
}+\hbar\omega_{q,\lambda}-\varepsilon_{k^{\prime}}.$ The satellite causes a
finite electron occupation of the states $\mathbf{k}^{\prime\prime}$ above the
Fermi surface and a finite hole occupation of the states $\mathbf{k}^{\prime}$
below the Fermi surface.

In Fig.1 an electron is introduced into a state $\mathbf{k}_{0}$ directly
above the Fermi energy. It changes the occupation of the $\mathbf{k}_{0}%
$-state only by $Z^{-1}<1$ because this state was already partially occupied,
and after the introduction into the state $\mathbf{k}_{0}$ part of the
electron makes virtual transitions into the other states above the Fermi
energy. This reduces the occupation of the state $\mathbf{k}_{0}$. Part of the
electron is smeared over an energy range of $\hbar\omega_{D}$ in combination
with single virtual phonons.

Therefore mass enhancement in an electron state $\mathbf{k}_{0}$ above the
Fermi energy has two contributions:

\begin{itemize}
\item The pre-occupation of the state $\mathbf{k}_{0}$ is larger than $0$: The
state $\mathbf{k}_{0}$ is already partially occupied before an additional
electron is introduced.

\item The post-occupation of the state $\mathbf{k}_{0}$ is less than $1$:
After the introduction of an additional electron the state is not completely
occupied because the electron makes virtual transitions into the other states
above the Fermi energy.
\end{itemize}

Both contributions are roughly equal.
\begin{align*}
&
{\includegraphics[
height=2.8717in,
width=3.8962in
]%
{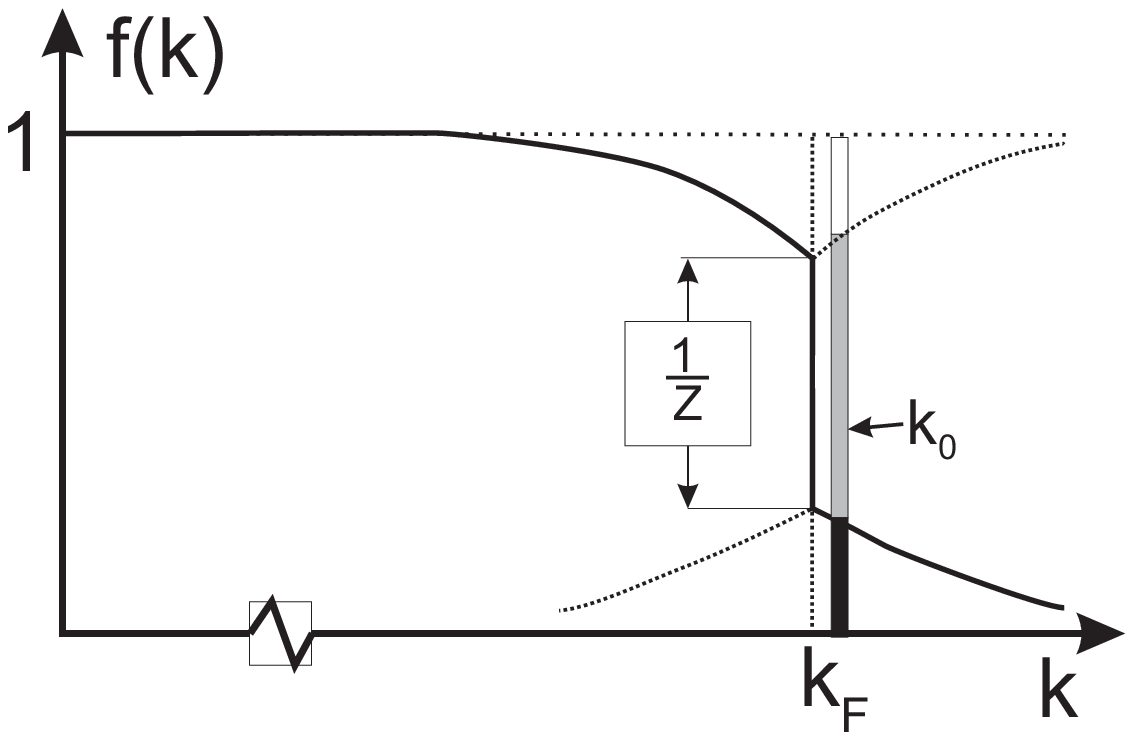}%
}%
\\
&
\begin{tabular}
[c]{l}%
Fig.1: The occupation as a function of momentum (energy) for\\
a strong-coupling metal. At the Fermi momentum (energy) the\\
jump in occupation at zero temperature is $1/Z$.
\end{tabular}
\end{align*}

Since the occupation of the state $\mathbf{k}_{0}$ changes only by $Z^{-1}$
the energy (of the quasi-particle in $\mathbf{k}_{0}$) is only $\varepsilon
_{k_{0}}/Z$. Therefore the quasi-particle energies are closer together and
their density of states is enhanced by the factor $Z$.

In appendix A1 a short review is given of the treatment of the electron-phonon
mass enhancement with Green functions. The discussed physical interpretation
of the mass enhancement in terms of pre- and post-occupation is sketched.
However, the Green function method cannot be easily expanded to multi-layers.
Therefore in this paper I use self-consistant perturbation theory.

In part II the electron-phonon interaction and matrix element in multi-layers
is derived. In part III the amplitudes of the electron-phonon satellites are
discussed. In a multi-layer these satellites interfere in real space. This is
discussed in part IV. In appendix A2 the ground-state wave function in the
presence of electron-phonon interaction is derived in self-consistent
perturbation theory. Finally appendix A3 treats the ground state plus an
additional electron above the Fermi energy.

Even in a homogeneous strong-coupling metal the derivation of the relation
$Z=1+\lambda$ is a complex and extended calculation. The multi-layer the
system is highly inhomogeneous, and I attempt a more modest goal. I calculate
the occupation of the electron states in real space in self-consistent
(Brillouin-Wigner) perturbation theory. This yields an expression for the
position and momentum dependent mass enhancement. A numerical evaluation is
left to future work.

Abbreviations: EPI = electron-phonon interaction, EPME = electron-phonon
matrix element.

\section{Electron-Phonon Interaction in Multi-Layers}

Now the question is: What happens in a double layer? Does the finite
occupation above the Fermi energy leak from the strong-coupling metal into the
weak-coupling metal? If this is the case then one finds an enhanced density of
states also in the normal metal (because an additional electron above the
Fermi energy does not change the occupation by one). To analyze this question
we investigate a very simple model of a double layer.

\subsection{Model}

\begin{itemize}
\item We consider a double layer of a strong-coupling metal S and a
weak-coupling or normal metal W. To keep the analysis simple a primitive cubic
lattice is used for both films with the same lattice constant $a$. The films
are parallel to the x-y-plane. The metal S consists of $N_{z}^{s}$ layers and
lies in the range $0\leq z\leq$ $d_{s}=N_{z}^{s}a,$ and the metal W has
$N_{z}^{w}$ layers in the regime $d_{s}\leq z\leq d_{s}+$ $d_{w}=L_{z}%
=N_{z}a,$ where $d_{w}=N_{z}^{w}a$ and $N_{z}$ is the total number of layers
in the z-direction. For normalization reasons the double film has a finite but
large extension in the x- and y-direction with the lengths $L_{x}=N_{x}a$ and
$L_{y}=N_{y}a$. Furthermore we use periodic boundary conditions in all three
dimensions. This represents a multi-layer of S and W. (We expect that this is
equivalent to an isolated double layer with half the thicknesses $d_{s}/2$ and
$d_{w}/2$). The total number of atoms is $N=N_{x}N_{y}\left(  N_{z}^{s}%
+N_{z}^{w}\right)  $.

\item The electron density in both films is the same and the electrons behave
as free electrons. The electron states are given by the wave number
$\mathbf{k}=\left(  k_{x},k_{y},k_{z}\right)  $ with the quantization
$k_{i}=\nu_{i}2\pi/L_{i}$, ($i=x,y,z$). (The quantization in the z-direction
extends over the total thickness $L_{z}=d_{s}+d_{w}$).

\item Furthermore we assume that the elastic properties of both metals are
identical. Therefore the phonons propagate through both films without
scattering. The same quantization as for electrons applies to the wave vector
$\mathbf{q}=\left(  q_{x},q_{y},q_{z}\right)  $ of the lattice oscillations
$q_{i}=\mu_{i}2\pi/L_{i}$, ($i=x,y,z$). Here $\mu_{i}$ lies in the range
$-N_{i}/2$%
$<$%
$\mu_{i}\leq N_{i}/2$.
\end{itemize}

Next we derive the electron-phonon matrix element (EPME) $g_{\mathbf{k}%
_{2}-\mathbf{k}_{1},\mathbf{q,}\lambda}$ for the transition of an electron
from a state $\mathbf{k}_{1}$ into $\mathbf{k}_{2}$ by absorbing a phonon
$\left(  \mathbf{q},\lambda\right)  $ or emitting a phonon $\left(
-\mathbf{q},\lambda\right)  $. The EPME is derived in a number of textbooks
\cite{G42}, \cite{S52}. Therefore only the essential results are given here.

The atoms oscillate with the amplitude $\mathbf{u}_{\mathbf{n}}$, and their
position is given by
\[
\mathbf{r}_{\mathbf{n}}=\mathbf{R}_{\mathbf{n}}+\mathbf{u}_{\mathbf{n}}%
\]
where $\mathbf{R}_{\mathbf{n}}=\mathbf{n}a$ is the average position of the
atom $\mathbf{n}$ with $\mathbf{n}\mathbf{=}\left(  n_{x},n_{y},n_{z}\right)
$ being integer.

In the lower strong-coupling film for $0\leq z\leq d_{s}$ the atoms have an
(somewhat artificial) electron potential
\[
U\left(  \mathbf{r}\right)  =V\left(  \mathbf{r-r}_{\mathbf{n}}\right)
-V\left(  \mathbf{r-R}_{\mathbf{n}}\right)
\]
In the upper weak-coupling film for $d_{s}\leq z\leq d_{s}+d_{w}$ the atoms
have zero electron potential. For zero displacement of the atoms the electron
potential vanishes in both films. Therefore we treat the electrons in both
films as free.

The potential due to the lattice oscillations is%

\[
U\left(  \mathbf{r}\right)  \thickapprox%
{\textstyle\sum_{\mathbf{n}_{s}}}
\left(  \mathbf{u}_{\mathbf{n}}\cdot\triangledown\right)  V\left(
\mathbf{r-R}_{\mathbf{n}}\right)  =i%
{\textstyle\sum_{\mathbf{n}_{s}}}
{\textstyle\sum_{\mathbf{p}}}
\left(  \mathbf{u}_{\mathbf{n}}\cdot\mathbf{p}\right)  V_{\mathbf{p}%
}e^{i\mathbf{p}\left(  \mathbf{r-R}_{\mathbf{n}}\right)  }%
\]
where the sum over $\mathbf{n}_{s}$ represents the sum over all atoms in S.
The displacement $\mathbf{u}_{\mathbf{n}}$ of the atom at $\mathbf{R}%
_{\mathbf{n}}$ is expressed in terms of the annihilation and creation
operators $a_{\mathbf{q}^{\prime}\mathbf{,\lambda}},a_{-\mathbf{q}^{\prime
}\mathbf{,\lambda}}^{\ast}$ for the phonons $\left(  \mathbf{q}^{\prime
}\mathbf{,\lambda}\right)  $ and $\left(  -\mathbf{q}^{\prime}\mathbf{,\lambda
}\right)  $.
\[
\mathbf{u}_{\mathbf{n}}=\frac{1}{\sqrt{N}}%
{\textstyle\sum_{\mathbf{q}^{\prime},\lambda}}
\left(  \frac{\hbar}{2M\omega_{\mathbf{q}^{\prime}\mathbf{,\lambda}}}\right)
^{1/2}\mathbf{e}_{\mathbf{q}^{\prime}\mathbf{,\lambda}}e^{i\mathbf{q}^{\prime
}\mathbf{R}_{\mathbf{n}}}\left(  a_{\mathbf{q}^{\prime}\mathbf{,\lambda}%
}+a_{-\mathbf{q}^{\prime}\mathbf{,\lambda}}^{\ast}\right)
\]
where $\mathbf{e}_{\mathbf{q}^{\prime},\lambda}$ is the unit vector of
polarization $\lambda$. This yields the electron-phonon interaction
hamiltonian%
\begin{equation}
H_{e-p}=%
{\textstyle\sum_{\mathbf{k}_{1}\mathbf{,k}_{2},\mathbf{q}^{\prime
}\mathbf{,\lambda}}}
g_{\mathbf{k}_{2}\mathbf{-k}_{1},\mathbf{q}^{\prime}\mathbf{,\lambda}%
}\mathbf{c}_{\mathbf{k}_{2}}^{\ast}c_{\mathbf{k}_{1}}\left(  a_{\mathbf{q}%
^{\prime}\mathbf{,\lambda}}+a_{-\mathbf{q}^{\prime}\mathbf{,\lambda}}^{\ast
}\right)  \label{Hep}%
\end{equation}
with the electron-phonon matrix element $g_{\mathbf{k}_{2}\mathbf{-k}%
_{1},\mathbf{q}^{\prime}\mathbf{,\lambda}}$
\begin{equation}
g_{\mathbf{k}_{2}\mathbf{-k}_{1},\mathbf{q}^{\prime}\mathbf{,\lambda}}%
=i\sqrt{N}\left(  \frac{\hbar}{2M\omega_{\mathbf{q}^{\prime}\mathbf{,\lambda}%
}}\right)  ^{1/2}\left(  \left(  \mathbf{k}_{2}\mathbf{-k}_{1}\right)
\cdot\mathbf{e}_{\mathbf{q}^{\prime}\mathbf{,\lambda}}\right)  V_{\mathbf{k}%
_{2}\mathbf{-k}_{1}}S_{f}\left(  \mathbf{q}^{\prime}\mathbf{-}\left(
\mathbf{k}_{2}\mathbf{-k}_{1}\right)  \right)  \label{epme}%
\end{equation}
where the structure factor $S_{f}\left(  \mathbf{Q}\right)  $ [with
$\mathbf{Q=q}^{\prime}\mathbf{-}\left(  \mathbf{k}_{2}\mathbf{-k}_{1}\right)
$] is given by%
\begin{align}
S_{f}\left(  \mathbf{Q}\right)   &  =\frac{1}{N}%
{\textstyle\sum_{\mathbf{n}_{s}}}
e^{i\mathbf{QR}_{\mathbf{n}}}=\frac{N_{x}N_{y}}{N}%
{\textstyle\sum_{G_{x},G_{y}}}
\delta_{Q_{x},G_{x}}\delta_{Q_{y},G_{y}}%
{\textstyle\sum_{n_{z}=0}^{N_{z}^{s}-1}}
e^{iQ_{z}n_{z}a}\label{StFaQ}\\
&  =%
{\textstyle\sum_{G_{x},G_{y}}}
\delta_{Q_{x},G_{x}}\delta_{Q_{y},G_{y}}\frac{1}{N_{z}}\frac{1-e^{iQ_{z}%
aN_{z}^{s}}}{1-e^{iQ_{z}a}}\nonumber
\end{align}
where $\mathbf{G}$ is a reciprocal lattice vector.

This means that there is no conservation of the z-component of the lattice
momentum for the electron-phonon processes. For comparison we denote the
electron-phonon matrix element for the pure metal S as $g_{\mathbf{k}%
_{2}\mathbf{-k}_{1},\mathbf{q}^{\prime}\mathbf{,\lambda}}^{0}$. Its structure
factor $S_{f}^{0}\left(  \mathbf{Q}\right)  =%
{\textstyle\sum_{\mathbf{G}}}
\delta_{\mathbf{Q,G}}$ with $\mathbf{Q=q}^{\prime}\mathbf{-}\left(
\mathbf{k}_{2}\mathbf{-k}_{1}\right)  $ fulfills conservation of lattice
momentum
\begin{equation}
g_{\mathbf{k}_{2}\mathbf{-k}_{1},\mathbf{q}^{\prime}\mathbf{,\lambda}}%
^{0}=i\sqrt{N}\left(  \frac{\hbar}{2M\omega_{\mathbf{q}^{\prime}%
\mathbf{,\lambda}}}\right)  ^{1/2}\left(  \left(  \mathbf{k}_{2}%
\mathbf{-k}_{1}\right)  \cdot\mathbf{e}_{\mathbf{q}^{\prime}\mathbf{,\lambda}%
}\right)  V_{\mathbf{k}_{2}\mathbf{-k}_{1}}%
{\textstyle\sum_{\mathbf{G}}}
\delta_{\mathbf{k}_{2}\mathbf{-k}_{1},\mathbf{q}^{\prime}+\mathbf{G}}
\label{epme0}%
\end{equation}

For those electron-phonon processes in the multi-layer which conserve the
lattice momentum, i.e., when $\mathbf{k}_{2}\mathbf{-k}_{1}\mathbf{=q}%
^{\prime}\mathbf{+G}$, one obtains%
\[
g_{\mathbf{k}_{2}\mathbf{-k}_{1},\mathbf{q}^{\prime}\mathbf{,\lambda}}%
=\frac{N_{z}^{s}}{N_{z}}g_{\mathbf{k}_{2}\mathbf{-k}_{1},\mathbf{q}^{\prime
}\mathbf{,\lambda}}^{0}=\frac{d_{s}}{d_{s}+d_{w}}g_{\mathbf{k}_{2}%
\mathbf{-k}_{1},\mathbf{q}^{\prime}\mathbf{,\lambda}}^{0}%
\]
The weight of the EPME which conserves lattice momentum is reduced by the
factor $d_{s}/\left(  d_{s}+d_{w}\right)  .$

\section{Electron-phonon satellites}

At $T=0$ in the absence of the electron-phonon interaction all states within
the Fermi sphere with $k\leq k_{F}$ ($k_{F}$ is the Fermi wave number) are
occupied and all other states are empty. We denote this state as $\left\vert
\Psi_{0}\right\rangle =%
{\textstyle\prod\limits_{k^{\prime}<k_{F}}}
c_{\mathbf{k}^{\prime}}^{\ast}\left\vert \Phi_{0}\right\rangle $ where
$\left\vert \Phi_{0}\right\rangle $ is the vacuum. We select from the full
Fermi sphere an occupied state $\mathbf{k}^{\prime}$. (In the following the
states such $\mathbf{k}$ will be sometimes denoted by their creation operators
such as $c_{\mathbf{k}}^{\ast}$). Furthermore we choose from the phonon
spectrum a phonon state $\left(  \mathbf{q},\lambda\right)  $. The state
$\mathbf{k}^{\prime}$ can make a transition into a state $\mathbf{k}%
^{\prime\prime}$ (above the Fermi surface) and create a phonon $\left(
\mathbf{q},\lambda\right)  $. Such a state can be described as a
$\mathbf{k}^{\prime\prime}$-electron - $\mathbf{k}^{\prime}$-hole plus one
phonon $\left(  \mathbf{q},\lambda\right)  $. We denote its amplitude as
$\alpha_{\mathbf{k}^{\prime\prime},\mathbf{k}^{\prime}\mathbf{,q}}$. The
resulting ground state $\widetilde{\left\vert \Psi_{0}\right\rangle }$ in the
presence of EPI is derived in the appendix A2 in self-consistent perturbation
theory. One obtains for the amplitude of the satellites%
\begin{equation}
\alpha_{\mathbf{k}^{\prime\prime},\mathbf{k}^{\prime}\mathbf{,q,}\lambda
}=\frac{g_{\mathbf{k}^{\prime\prime}-\mathbf{k}^{\prime},-\mathbf{q,}\lambda}%
}{\left(  \eta_{\mathbf{k}^{\prime}}^{0}-\varepsilon_{\mathbf{k}^{\prime
\prime}}-\hbar\omega\right)  }\label{a_k''}%
\end{equation}
where $\eta_{\mathbf{k}^{\prime}}^{0}$ is given in appendix A2 by the
self-consistent equation (\ref{eth}).

This means that a state $c_{\mathbf{k}^{\prime}}^{\ast}$ with the occupation
"$1$" generates satellites $c_{\mathbf{k}^{\prime\prime}}^{\ast}%
a_{\mathbf{q},\lambda}^{\ast}$ with the (relative) occupation $\left\vert
\alpha_{\mathbf{k}^{\prime\prime},\mathbf{k}^{\prime}\mathbf{,q,}\lambda
}\right\vert ^{2}$. After normalizing the total electron state one obtains for
the occupation of an electron state $\mathbf{k}_{0}^{\prime}$ below the Fermi
energy and an electron $\mathbf{k}_{0}^{\prime\prime}$ above the Fermi energy
\begin{align}
n\left(  \mathbf{k}_{0}^{\prime}\right)   &  =\frac{1}{1+\left(
{\textstyle\sum_{\mathbf{k}^{\prime\prime},\mathbf{q},\lambda}}
\left\vert \alpha_{\mathbf{k}^{\prime\prime},\mathbf{k}_{0}^{\prime
}\mathbf{,q,}\lambda}\right\vert ^{2}\right)  }\label{n_oc}\\
n\left(  \mathbf{k}_{0}^{\prime\prime}\right)   &  =\frac{%
{\textstyle\sum_{\mathbf{k}^{\prime},\mathbf{q},\lambda}}
\left\vert \alpha_{\mathbf{k}_{0}^{\prime\prime},\mathbf{k}^{\prime
}\mathbf{,q,}\lambda}\right\vert ^{2}}{1+\left(
{\textstyle\sum_{\mathbf{k}^{\prime},\mathbf{q},\lambda}}
\left\vert \alpha_{\mathbf{k}_{0}^{\prime\prime},\mathbf{k}^{\prime
}\mathbf{,q,}\lambda}\right\vert ^{2}\right)  }\nonumber
\end{align}

The step at the Fermi energy is reduced, because (i) a state $\mathbf{k}%
_{0}^{\prime\prime}$ above the Fermi energy is partially occupied and (ii) a
state $\mathbf{k}_{0}^{\prime}$ below the Fermi energy is partially empty. As
we discussed above both effects contribute to the electron mass enhancement.

\section{The interference of the satellite wave functions in a multi-layer}

In a multi-layer one expects that the occupations $n\left(  \mathbf{k}%
_{0}^{\prime}\right)  $ and $n\left(  \mathbf{k}_{0}^{\prime\prime}\right)  $
in the ground state for energies below and above the Fermi energy are position
dependent. For simplicity we ignore umklapp processes with $\mathbf{G}\neq0.$
Then, in a homogeneous metal, one obtains for a given $\mathbf{k}^{\prime}%
$-state within the Fermi sphere and a given phonon state $\left(
\mathbf{q,}\lambda\right)  $ just one satellite state $c_{\mathbf{k}%
^{\prime\prime}}^{\ast}a_{\mathbf{q},\lambda}^{\ast}$ with $\mathbf{k}%
^{\prime\prime}=\mathbf{k}^{\prime}\mathbf{-q}$. This is different for the
double layer. Here $\mathbf{k}^{\prime\prime}$ can take many possible values,%
\[
\mathbf{k}^{\prime\prime}=\mathbf{k}^{\prime}\mathbf{-q}+\nu\frac{2\pi}%
{d_{s}+d_{w}}\widehat{\mathbf{z}}=\mathbf{k}_{0}^{\prime\prime}+\nu\mathbf{g}%
\]
where $\mathbf{k}_{0}^{\prime\prime}=\mathbf{k}^{\prime}-\mathbf{q}$ and
$\mathbf{g=}\frac{2\pi}{d_{s}+d_{w}}\widehat{\mathbf{z}}.$ We call the states
$c_{\mathbf{k}_{0}^{\prime\prime}\mathbf{+\nu g}}^{\ast}a_{\mathbf{q},\lambda
}^{\ast}$ with the amplitude $g_{\mathbf{-q}+\nu\mathbf{g},-\mathbf{q,}%
\lambda}/\left(  \eta_{\mathbf{k}^{\prime}}-\varepsilon_{\mathbf{k}%
_{0}^{\prime\prime}+\nu\mathbf{g}}-\hbar\omega\right)  $ a family of satellite
states. Each member of the family has the same electron hole $\mathbf{k}%
^{\prime}$and the same phonon $\left(  \mathbf{q},\lambda\right)  $. The
members only differ in the wave number of the state $\mathbf{k}^{\prime\prime
}$ by $\nu\mathbf{g}$. The states of a family are coherent and can interfere.
Their phases differ by $\exp\left(  i\nu gz\right)  $. This yields a
modulation of the amplitude of $\left\vert c_{\mathbf{k}_{0}^{\prime\prime}%
}^{\ast}c_{\mathbf{k}^{\prime}}a_{\mathbf{q},\lambda}^{\ast}\Psi
_{0}\right\rangle $ in real space.

The total amplitude of the family $\left\vert c_{\mathbf{k}_{0}^{\prime\prime
}}^{\ast}c_{\mathbf{k}^{\prime}}a_{\mathbf{q},\lambda}^{\ast}\Psi
_{0}\right\rangle $ of hole-electron-phonon states in real space is%
\[
A_{\mathbf{k}_{0}^{\prime\prime}\mathbf{,k}^{\prime},\mathbf{q},\lambda
}\left(  \mathbf{r}\right)  =%
{\textstyle\sum_{\nu}}
a_{\mathbf{k}_{0}^{\prime\prime}\mathbf{,k}^{\prime},\mathbf{q},\lambda
}e^{i\nu\mathbf{gr}}=%
{\textstyle\sum_{\nu}}
\frac{g_{\mathbf{-q}+\nu\mathbf{g},-\mathbf{q,}\lambda}}{\left(
\eta_{\mathbf{k}^{\prime}}-\varepsilon_{\mathbf{k}_{0}^{\prime\prime}%
+\nu\mathbf{g}}-\hbar\omega\right)  }e^{i\nu\mathbf{gr}}%
\]
With $Q_{z}=-q_{z}-\left(  k_{z}^{\prime\prime}-k_{z}^{\prime}\right)
=\allowbreak-g\nu$ we obtain with equ.(\ref{epme})
\begin{equation}
=i\sqrt{N}\left(  \frac{\hbar}{2M\omega_{q\mathbf{,\lambda}}}\right)  ^{1/2}%
{\textstyle\sum_{\nu}}
\frac{\left(  \left(  -\mathbf{q+\nu g}\right)  \mathbf{\cdot e}%
_{\mathbf{q,\lambda}}\right)  V_{-\mathbf{q+\nu g}}}{\left(  \eta
_{\mathbf{k}^{\prime}}-\varepsilon_{\mathbf{k}_{0}^{\prime\prime}%
+\nu\mathbf{g}}-\hbar\omega\right)  }\frac{1}{N_{z}}\frac{1-\exp\left(  -i\nu
gN_{z}^{s}a\right)  }{1-\exp\left(  -i\nu ga\right)  }e^{i\mathbf{\nu}gz}
\label{Ak0}%
\end{equation}

Before we evaluate this result in more detail we consider two extreme cases:

\begin{enumerate}
\item Only the term with $\nu=0$ contributes. Then we have%
\begin{align*}
A_{\mathbf{k}_{0}^{\prime\prime}\mathbf{,k}^{\prime},\mathbf{q},\lambda
}\left(  \mathbf{r}\right)   &  =i\sqrt{N}\left(  \frac{\hbar}{2M\omega
_{q\mathbf{,\lambda}}}\right)  ^{1/2}\frac{\left(  -\mathbf{q\cdot
e}_{\mathbf{q,\lambda}}\right)  V_{-\mathbf{q}}}{\left(  \eta_{\mathbf{k}%
^{\prime}}-\varepsilon_{\mathbf{k}_{0}^{\prime\prime}}-\hbar\omega\right)
}\frac{N_{z}^{s}}{N_{z}}\\
&  =\frac{N_{z}^{s}}{N_{z}}\frac{g_{\mathbf{-q},-\mathbf{q,}\lambda}^{0}%
}{\left(  \eta_{\mathbf{k}^{\prime}}-\varepsilon_{\mathbf{k}_{0}^{\prime
\prime}}-\hbar\omega\right)  }%
\end{align*}
where $g_{\mathbf{-q},-\mathbf{q,}\lambda}^{0}$ is the electron-phonon matrix
element for the homogeneous metal S. This wave function has constant density
in the metals S and W. The electron-phonon matrix element of the metal S is
averaged over the total double-layer thickness and reduced by the factor of
$d_{s}/\left(  d_{s}+d_{w}\right)  $. The satellite state is equally
distributed over both films and partially blocks the state $\mathbf{k}%
_{0}^{\prime\prime}$ in both films equally. This means that both films have an
identical enhancement factor.

\item If the contribution of the term $\nu\mathbf{g}$ can be neglected in the
expressions $\left(  \eta_{\mathbf{k}^{\prime}}-\varepsilon_{\mathbf{k}%
_{0}^{\prime\prime}+\nu\mathbf{g}}-\hbar\omega\right)  $ and $\left(  \left(
-\mathbf{q+\nu g}\right)  \mathbf{\cdot e}_{\mathbf{q,\lambda}}\right)
V_{-\mathbf{q+\nu g}}$ then we obtain
\begin{align*}
A_{\mathbf{k}^{\prime}\mathbf{-q,k}^{\prime},\mathbf{q},\lambda}\left(
\mathbf{r}\right)   &  =i\sqrt{N}\left(  \frac{\hbar}{2M\omega
_{q\mathbf{,\lambda}}}\right)  ^{1/2}\frac{\left(  -\mathbf{q\cdot
e}_{\mathbf{q,\lambda}}\right)  V_{-\mathbf{q}}}{\left(  \eta_{\mathbf{k}%
^{\prime}}-\varepsilon_{\mathbf{k}_{0}^{\prime\prime}}-\hbar\omega\right)  }\\
&  \ast%
{\textstyle\sum_{\nu=-N_{z}/2+1}^{N_{z}/2}}
\frac{1}{N_{z}}\frac{1-\exp\left(  -\frac{2\pi i\nu N_{z}^{s}}{N_{z}}\right)
}{1-\exp\left(  -\frac{2\pi\nu i}{N_{z}}\right)  }\exp\left(  \frac{2\pi i\nu
}{L_{z}}z\right)
\end{align*}
This yields%
\begin{align*}
A_{\mathbf{k}^{\prime}\mathbf{-q,k}^{\prime},\mathbf{q},\lambda}\left(
\mathbf{r}\right)   &  =\frac{g_{\mathbf{-q},-\mathbf{q,}\lambda}^{0}}{\left(
\eta_{\mathbf{k}^{\prime}}-\varepsilon_{\mathbf{k}_{0}^{\prime\prime}}%
-\hbar\omega\right)  }S\left(  z\right)  \\
S\left(  z\right)   &  =%
{\textstyle\sum_{\nu=-N_{z}/2+1}^{N_{z}/2}}
\frac{1}{N_{z}}\frac{1-\exp\left(  -\frac{2\pi i\nu N_{z}^{s}}{N_{z}}\right)
}{1-\exp\left(  -\frac{2\pi\nu i}{N_{z}}\right)  }\exp\left(  \frac{2\pi i\nu
}{L_{z}}z\right)
\end{align*}
where $S\left(  z\right)  $ is essentially a step function which is equal to
one in the strong-coupling metal S and zero in the normal metal W. In Fig.2
the function $S\left(  z/a\right)  $ is shown for a multi-layer with
$N_{z}^{s}=10$ and $N_{z}^{w}=8$. If the conditions for case (2) are fulfilled
the state $\mathbf{k}^{\prime\prime}$ has its full amplitude in the metal S
and its amplitude is essentially zero in the normal metal W. In this case we
expect the full mass enhancement in S and no mass enhancement in W.
\begin{align*}
&
{\includegraphics[
height=2.3645in,
width=3.0178in
]%
{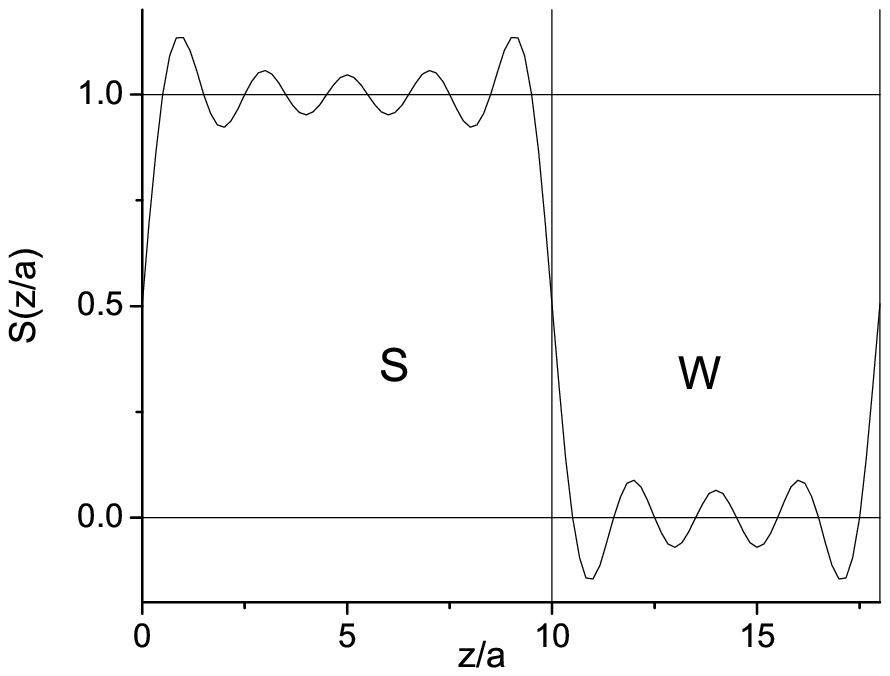}%
}%
\\
&
\begin{tabular}
[c]{l}%
Fig.2: The amplitude of the electronic part of the\\
electron-phonon satellite (with the wave function $e^{-i\mathbf{k}_{0}%
^{\prime\prime}\mathbf{r}}$)\\
is restricted to the strong-coupling metal S and vanishes\\
in the normal metal W.
\end{tabular}
\end{align*}

\end{enumerate}

For a realistic evaluation we consider a sandwich of 10 atomic layers of S and
8 atomic layers of W. For S and W we use the electronic density and Debye
temperature of Pb with $\varepsilon_{F}=9.5eV,$ $k_{F}=1.6\times10^{10}m^{-1}$
and $\Theta_{D}=90K$ but assume a simple cubic lattice with $a=3\allowbreak
.28\times10^{-10}m$. The vector $g$ has the value of $1.\,\allowbreak
9\times10^{9}m^{-1}$.

Fig.3 shows a typical electron-phonon process. The electron in the state
$\mathbf{k}^{\prime}$ below the Fermi surface emits a phonon $\left(
\mathbf{q},\lambda\right)  $ and makes a transition into a state
$\mathbf{k}^{\prime\prime}$ above the Fermi surface. The amplitude in the
satellite state is proportional to the inverse energy denominator $\left(
\eta_{\mathbf{k}^{\prime}}-\varepsilon_{\mathbf{k}^{\prime\prime}}-\hbar
\omega\right)  ^{-1}$. Therefore the main contribution is from the regime were
$\left\vert \varepsilon_{\mathbf{k}^{\prime\prime}}\right\vert ,\left\vert
\varepsilon_{\mathbf{k}^{\prime}}\right\vert <<\hbar\omega_{D}$ and the states
$\mathbf{k}^{\prime}$ and $\mathbf{k}^{\prime\prime}$ lie close the Fermi energy.%

\begin{align*}
&
{\includegraphics[
height=4.0672in,
width=3.5259in
]%
{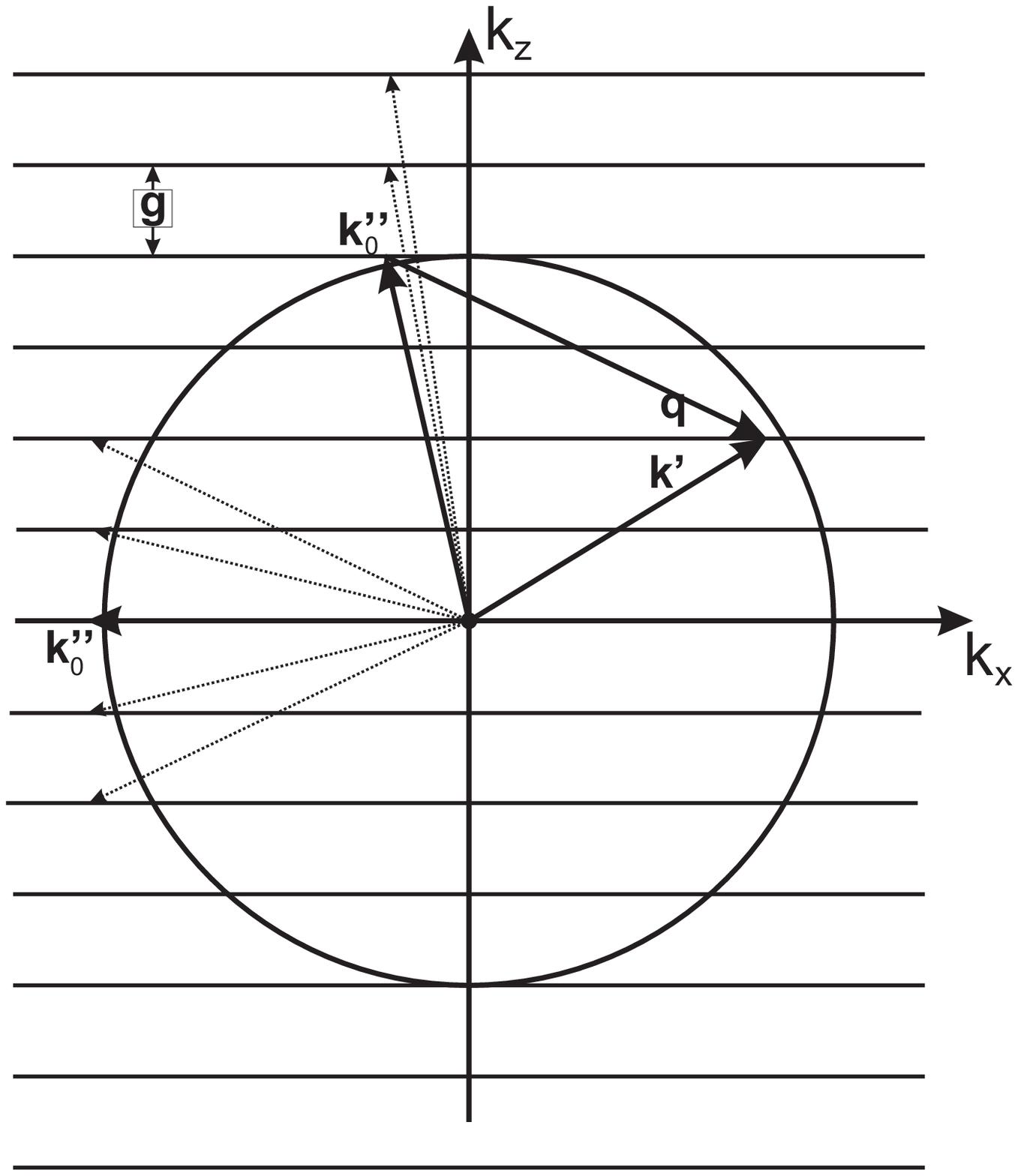}%
}%
\\
&
\begin{tabular}
[c]{l}%
Fig.3: The $k_{z}$-planes of a double layer with periodic boundary\\
conditions. Both metals have the same electron band structure.\\
The films have strong and zero electron-phonon interaction.\\
The arrows show transitions from state $\mathbf{k}^{\prime}$ to $\mathbf{k}%
^{\prime\prime}$ with a virtual\\
phonon $\left(  \mathbf{q},\lambda\right)  $. For given $\mathbf{k}^{\prime
}\mathbf{,q}$ there are several final states\ \\
$\mathbf{k}^{\prime\prime}=\mathbf{k}^{\prime}\mathbf{-q+}\nu2\pi
\widehat{\mathbf{z}}\mathbf{/}\left(  d_{s}+d_{w}\right)  $ permitted. The
amplitude of these\\
states are coherent and interfere.
\end{tabular}
\end{align*}

In one scenario the main satellite state $\mathbf{k}_{0}^{\prime\prime}$ lies
close to the z-direction. In this case the sub-states $\mathbf{k}%
^{\prime\prime}=\mathbf{k}_{0}^{\prime\prime}+\nu\mathbf{g}$ with negative
$\nu$ are occupied and not available. The sub-states with positive $\nu$ lie
above the Fermi energy by an energy of $\delta E\left(  \nu\right)
=\nu\varepsilon_{F}\frac{g}{2k_{F}}\thickapprox0.06\nu\varepsilon_{F}$. Since
the Fermi energy corresponds to a temperature of about $1.1\times10^{5}K$ the
next sub-state $\mathbf{k}^{\prime\prime}=\mathbf{k}_{0}^{\prime\prime
}+\mathbf{g}$ lies above the Fermi level by an energy corresponding to
$6600K$. This is very large compared to the Debye temperature of $90K.$
Because of the large energy denominator $\left(  \eta_{\mathbf{k}^{\prime}%
}-\varepsilon_{\mathbf{k}_{0}^{\prime\prime}}-\hbar\omega\right)  ^{-1}$ the
sub-states with $\mathbf{k}^{\prime\prime}=\mathbf{k}_{0}^{\prime\prime}%
+\nu\mathbf{g}$ can be neglected. In this direction only the main state
$\mathbf{k}_{0}^{\prime\prime}$ contributes, i.e., only the value $\nu=0$
contributes as discussed in case (1). It is remarkable that one can increase
the film thicknesses by a factor of 100 before the energy $\delta E\left(
\nu=1\right)  $ is of the order of the Debye energy.

The situation is different when $\mathbf{k}_{0}^{\prime\prime}$ lies in the
x-direction. This is shown in Fig.3 on the left side where $\mathbf{k}%
_{0}^{\prime\prime}$ point in the negative x-direction. Here the states
$\mathbf{k}^{\prime\prime}=\mathbf{k}_{0}^{\prime\prime}+\nu\mathbf{g}$ are
available. Their energy separation from the state $\mathbf{k}_{0}%
^{\prime\prime}$ is given by $\delta E\left(  \nu\right)  $ $=\left(  \hbar\nu
g\right)  ^{2}/\left(  2m\right)  =$ $\left(  \nu g/k_{F}\right)
^{2}\varepsilon_{F}$ $\thickapprox0.014\nu^{2}\varepsilon_{F}$ \ $\thickapprox
\nu^{2}1540K$. Again this value lies considerably above the Debye temperature
of $90K$. However when we increase the thicknesses $d_{s}$ and $d_{w}$ by a
factor 10 then $\delta E$ reduces to $15K$ and the sub-satellites have to be
included in the calculation.

This background occupation (at $T=0$) yields about one half of the mass
enhancement. From the dependence of the energy separation $\delta E\left(
\nu\right)  $ on $\nu$ and $g$ we obtain the following results as a function
of the total thickness $\left(  d_{s}+d_{w}\right)  =N_{z}a$:

\begin{itemize}
\item $N_{z}<50$: The electron-phonon matrix element is reduced by the factor
$p=N_{z}^{s}/\left(  N_{z}^{s}+N_{z}^{w}\right)  $. Both films are equally
enhanced but the enhancement factor is reduced by $p^{2}$.

\item $50<N_{z}<300$: The occupation of a state $\mathbf{k}_{0}^{\prime\prime
}$ depends critically on the direction of $\mathbf{k}_{0}^{\prime\prime}$; for
$\mathbf{k}_{0}^{\prime\prime}$ parallel to the z-direction its occupation is
the same in S and W. Here one still has an averaged EPME. For $\mathbf{k}%
_{0}^{\prime\prime}$ parallel to the film plane the occupation of
$\mathbf{k}_{0}^{\prime\prime}$ in S takes its full value while it approaches
zero in W. The enhanced density of states in both films is highly anisotropic.

\item $N_{z}>300$: The occupation of the state $\mathbf{k}_{0}^{\prime\prime}$
approaches the individual value for the two metals S and W.
\end{itemize}

The second half of the mass enhancement is due to virtual electron-phonon
processes which start from the quasi-particle state $\mathbf{k}_{0}$ with
final states $\mathbf{k}^{\prime\prime},\left(  \mathbf{q},\lambda\right)  $
where the sum goes over all free electron states $\mathbf{k}^{\prime\prime}$
above the Fermi energy. As before the weight of these processes in S and W
depends strongly on the final state. However, since now we have to sum over
the final states $\mathbf{k}^{\prime\prime}$ the resulting anisotropy for the
state $\mathbf{k}_{0}$ is strongly reduced. One obtains as before the thin
film regime for $N_{z}<50$ where the EPME is averaged over the two metals and
the bulk limit for $N_{a}>300$ where the individual bulk mass enhancement are
reinstated. In the intermediate thickness range one has a slow transition
between the two extremes.

\section{Conclusions}

In this paper mass enhancement is investigated in double and multi-layers of
two metals with strong and weak-coupling electron-phonon interaction. The mass
enhancement is due to the fact that an electron injected into a state
$\mathbf{k}_{0}$ above the Fermi energy changes the occupation of this state
by less than one. This is for two reasons: (i) the state $\mathbf{k}_{0}$ was
(even at $T=0$) already partially occupied due to electron transitions from
the occupied Fermi sea into the state $\mathbf{k}_{0}$ emitting a virtual
phonon $\left(  \mathbf{q},\lambda\right)  $ and (ii) the injected electron
makes transitions into states $\mathbf{k}^{\prime\prime}$ above the Fermi
surface emitting virtual phonons $\left(  \mathbf{q},\lambda\right)  $ and
reducing the occupation of the state $\mathbf{k}_{0}$. In multi-layers one has
modified electron-phonon matrix elements and a finite quantization of electron
and phonon states perpendicular to the film planes. As a consequence one has
to consider interference between electron-phonon processes which start from
the same initial electron state $\mathbf{k}^{\prime}$ ($k^{\prime}<k_{F}$),
emit the same phonon $\left(  \mathbf{q},\lambda\right)  $ and yield a
superposition of different $c_{\mathbf{k}^{\prime\prime}}^{\ast}$ for the
final electron state. After the emission of the virtual phonon the electron no
longer has a well defined momentum. This interference yields a spatial
dependence of the pre-occupation of the electron state $\mathbf{k}_{0}$. This
results in very interesting properties of the mass enhancement in multi-layers
of metals with strong and weak-coupling electron-phonon interaction. In the
thin-film limit the electron-phonon matrix element is averaged over both
films. In an intermediate thickness range 50\% of the mass enhancement in each
film depends strongly on the direction of the electron momentum $\mathbf{k}%
_{0}$. In both films the mass enhancement approaches the bulk value in the
direction parallel to the film planes while perpendicular to the films one
obtains an averaged mass enhancement. This will cause a rather anisotropic
propagation of the conduction electrons parallel and perpendicular to the
films. In strong-coupling superconductors in contact with normal films it will
influence the boundary condition between the films and as a consequence the
superconducting transition temperature of the double or multi layer as well
the upper critical field. Even for the simple model which is considered in
this paper an extensive numerical calculation is required to obtain the
details of the mass enhancement because it depends on the direction of the
electron wave number $\mathbf{k}_{0}$ in both metals.

\newpage

\appendix{}

\section{Appendix}

\subsection{Connection with Green function self energy}

In the derivation of the mass enhancement in a pure bulk metal S one generally
starts from the fully occupied free electron Fermi sphere. A standard
treatment uses the Green-function method. Here one calculates the self-energy
of an additional electron $\mathbf{k}_{0}$ just above the Fermi energy (see
for example \cite{S52}, \cite{S68}, \cite{G42}). This has two contributions.
Fig.4 shows the well-known processes involved. \newline(a) The inserted
electron in state $\mathbf{k}_{0}$ emits a phonon $\left(  \mathbf{q}%
,\lambda\right)  $ and makes a virtual transition into the state
$\mathbf{k}^{\prime\prime}$.\newline(b) An electron in the state
$\mathbf{k}_{0}$ blocks all electron-phonon processes in which an electron
from an occupied state $\mathbf{k}^{\prime}$ emits a phonon $\left(
\mathbf{q},\lambda\right)  $ and makes a transition into the state
$\mathbf{k}_{0}$. \newline%

\begin{align*}
&
{\includegraphics[
height=2.4931in,
width=4.0498in
]%
{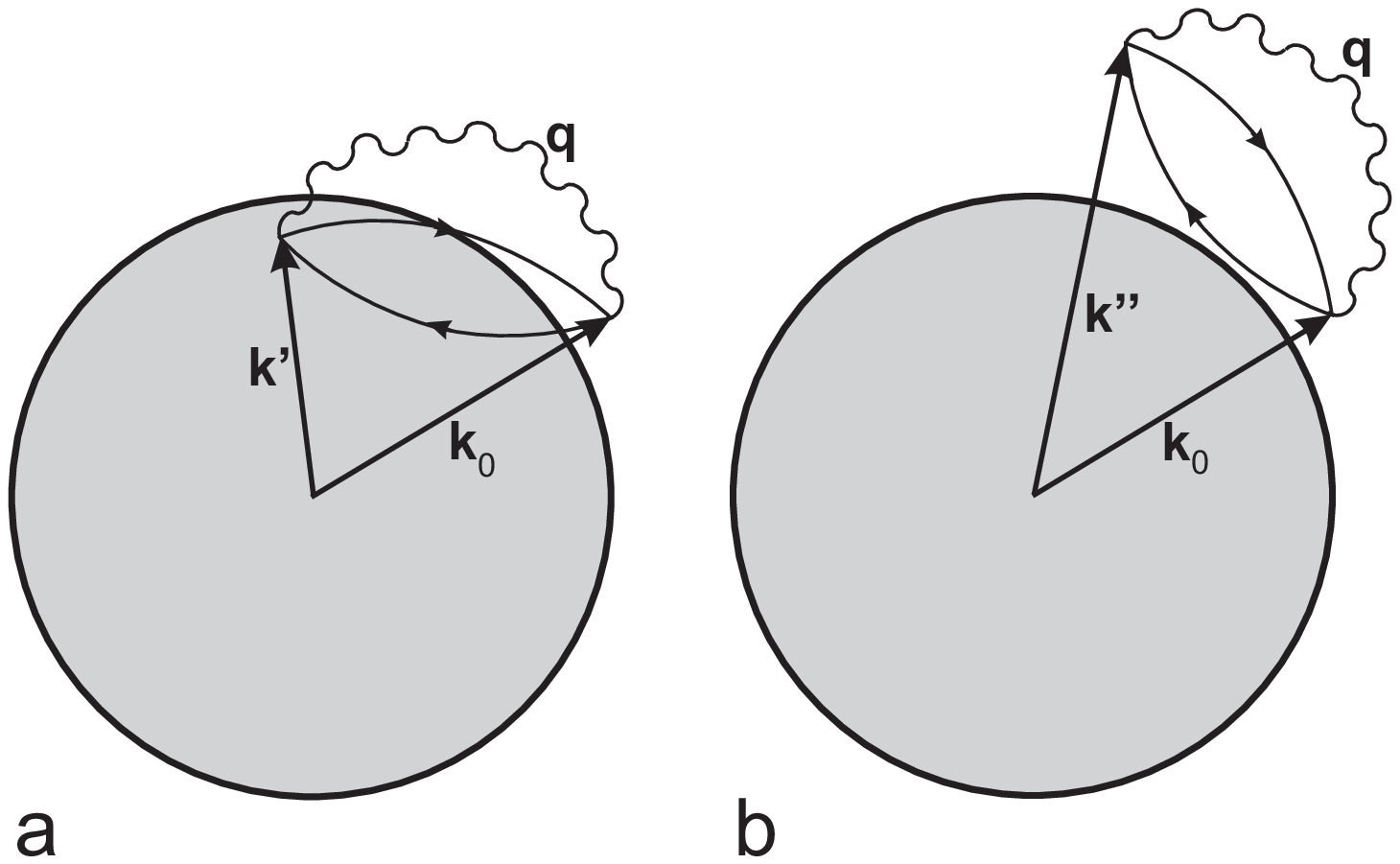}%
}%
\\
&
\begin{tabular}
[c]{l}%
Fig.4: The two contributions to the self energy and\\
the mass enhancement
\end{tabular}
\end{align*}
The corresponding self-energies of the state $\mathbf{k}^{\prime\prime}$ are
\begin{align*}
\Sigma_{a}\left(  \mathbf{k}_{0},E\right)   &  =%
{\textstyle\sum_{\mathbf{k}^{\prime\prime}\mathbf{,q,}\lambda}}
\frac{\left\vert g_{\mathbf{k}^{\prime\prime}-\mathbf{k}_{0}\mathbf{,-q,}%
\lambda}\right\vert ^{2}}{E-\varepsilon_{k^{\prime\prime}}-\hbar\omega
+i\delta}\\
\Sigma_{b}\left(  \mathbf{k}_{0},E\right)   &  =-%
{\textstyle\sum_{\mathbf{k}^{\prime}\mathbf{,q,}\lambda}}
\frac{\left\vert g_{\mathbf{k}_{0}-\mathbf{k}^{\prime}\mathbf{,-q,}\lambda
}\right\vert ^{2}}{E-\varepsilon_{\mathbf{k}^{\prime}}+\hbar\omega+i\delta}%
\end{align*}

The inverse dressed Green function has then the form%
\[
G^{-1}=E-\varepsilon_{\mathbf{k}_{0}}+i\delta-\Sigma
\]
where $\Sigma=\Sigma_{a}+\Sigma_{b}$. The new quasi-particle energy is given
by the pole of $G$, i.e. by solving the implicit equation%

\begin{align}
E_{\mathbf{k}_{0}}  &  =\varepsilon_{\mathbf{k}_{0}}+\Sigma\left(
\mathbf{k}_{0},E_{\mathbf{k}_{0}}\right) \label{Ek0GF}\\
&  =\varepsilon_{\mathbf{k}_{0}}+%
{\textstyle\sum_{\mathbf{k}^{\prime\prime}\mathbf{,q,}\lambda}}
\frac{\left\vert g_{\mathbf{k}^{\prime\prime}-\mathbf{k}_{0}\mathbf{,-q,}%
\lambda}\right\vert ^{2}}{E_{\mathbf{k}_{0}}-\varepsilon_{k^{\prime\prime}%
}-\hbar\omega+i\delta}-%
{\textstyle\sum_{\mathbf{k}^{\prime}\mathbf{,q,}\lambda}}
\frac{\left\vert g_{\mathbf{k}_{0}-\mathbf{k}^{\prime}\mathbf{,-q,}\lambda
}\right\vert ^{2}}{E_{\mathbf{k}_{0}}-\varepsilon_{\mathbf{k}^{\prime}}%
+\hbar\omega+i\delta}\nonumber
\end{align}

The self-energy depends on the momentum $\mathbf{k}_{0}$ only through the
matrix element, and this dependence is very weak and can be neglected. If
$\mathbf{k}_{0}$ lies directly above the Fermi energy, i.e. $\mathbf{k}%
_{0}=\mathbf{k}_{F}^{+}$ then we have the relation%
\[
E_{\mathbf{k}_{F}^{+}}=\Sigma\left(  \mathbf{k}_{F}^{+},E_{\mathbf{k}_{F}^{+}%
}\right)
\]

Next we can expand $\Sigma\left(  \mathbf{k}_{0},E\right)  $ in terms of $E$
about the energy $E_{\mathbf{k}_{F}^{+}}$
\begin{align*}
\Sigma\left(  \mathbf{k}_{0},E\right)   &  =\Sigma\left(  \mathbf{k}_{F}%
^{+},E_{\mathbf{k}_{F}^{+}}\right)  +\left(  E-E_{\mathbf{k}_{F}^{+}}\right)
\frac{\partial}{\partial E}\Sigma\left(  \mathbf{k}_{F}^{+},E_{\mathbf{k}%
_{F}^{+}}\right)  \\
&  =\Sigma\left(  \mathbf{k}_{F}^{+},E_{\mathbf{k}_{F}^{+}}\right)
-E_{\mathbf{k}_{F}^{+}}\frac{\partial}{\partial E}\Sigma\left(  \mathbf{k}%
_{F}^{+},E_{\mathbf{k}_{F}^{+}}\right)  +E\frac{\partial}{\partial E}%
\Sigma\left(  \mathbf{k}_{F}^{+},E_{\mathbf{k}_{F}^{+}}\right)
\end{align*}
This yields for for the quasi-particle energy
\[
E_{\mathbf{k}_{0}}=\varepsilon_{\mathbf{k}_{0}}+\Sigma\left(  \mathbf{k}%
_{F}^{+},E_{\mathbf{k}_{F}^{+}}\right)  +\left(  E_{\mathbf{k}_{0}%
}-E_{\mathbf{k}_{F}^{+}}\right)  \frac{\partial}{\partial E}\operatorname{Re}%
\Sigma\left(  \mathbf{k}_{F}^{+},E_{\mathbf{k}_{F}^{+}}\right)
\]%
\[
E_{\mathbf{k}_{0}}=\frac{1}{\left(  1-\frac{\partial}{\partial E}%
\operatorname{Re}\Sigma\left(  \mathbf{k}_{F}^{+},E_{\mathbf{k}_{F}^{+}%
}\right)  \right)  }\left[  \varepsilon_{\mathbf{k}_{0}}+\Sigma\left(
\mathbf{k}_{F}^{+},E_{\mathbf{k}_{F}^{+}}\right)  -E_{\mathbf{k}_{F}^{+}}%
\frac{\partial}{\partial E}\Sigma\left(  \mathbf{k}_{F}^{+},E_{\mathbf{k}%
_{F}^{+}}\right)  \right]
\]
The terms $-\operatorname{Re}\Sigma\left(  \mathbf{k}_{F}^{+},E_{\mathbf{k}%
_{F}^{+}}\right)  +E_{\mathbf{k}_{F}^{+}}\frac{\partial}{\partial
E}\operatorname{Re}\Sigma\left(  \mathbf{k}_{F}^{+},E_{\mathbf{k}_{F}^{+}%
}\right)  $ yield essentially a constant energy shift which will be absorbed
in the chemical potential. Then the Green function takes the form%
\[
G\left(  \mathbf{k}_{0},E\right)  =\frac{1}{E-\varepsilon_{\mathbf{k}_{0}%
}-E\frac{\partial}{\partial E}\operatorname{Re}\Sigma\left(  \mathbf{k}%
_{F}^{+},E_{\mathbf{k}_{F}^{+}}\right)  +i\Gamma_{\mathbf{k}_{0}}}%
\]
or%
\[
G\left(  \mathbf{k}_{0},E\right)  =\frac{1}{Z}\frac{1}{E-\frac{\varepsilon
_{\mathbf{k}_{0}}}{Z}+i\frac{\Gamma_{\mathbf{k}_{0}}}{Z}}%
\]
where
\[
Z=\left(  1-\frac{\partial}{\partial E}\Sigma\left(  \mathbf{k}_{F}%
^{+},E_{\mathbf{k}_{F}^{+}}\right)  \right)  >1
\]
and $\Gamma_{\mathbf{k}_{0}}$ is an imaginary contribution from the self
energy which we neglected in the discussion.

When one performs the derivative of the self energy one realizes that its real
part represents the realative occupation of the states $\mathbf{k}%
^{\prime\prime}$. As an example we take $\Sigma_{a}$ and obtain for
$-\partial\operatorname{Re}\Sigma_{a}\left(  \mathbf{k}_{F}^{+},E_{\mathbf{k}%
_{F}^{+}}\right)  /\partial E$.
\[
-\frac{\partial}{\partial E}%
{\textstyle\sum_{\mathbf{k}^{\prime\prime}\mathbf{,q,}\lambda}}
\frac{\left\vert g_{\mathbf{k}^{\prime\prime}-\mathbf{k}_{0}\mathbf{,-q,}%
\lambda}\right\vert ^{2}}{E-\varepsilon_{k^{\prime\prime}}-\hbar\omega
}|_{E_{\mathbf{k}_{F}}}=%
{\textstyle\sum_{\mathbf{k}^{\prime\prime}\mathbf{,q,}\lambda}}
\frac{\left\vert g_{\mathbf{k}^{\prime\prime}-\mathbf{k}_{0}\mathbf{,-q,}%
\lambda}\right\vert ^{2}}{\left(  E_{\mathbf{k}_{F}^{+}}-\varepsilon
_{k^{\prime\prime}}-\hbar\omega\right)  ^{2}}%
\]
For each $\mathbf{k}^{\prime\prime},\mathbf{q}$ the right sum represents the
occupation of the electron state $\mathbf{k}^{\prime\prime}$ and a phonon
$\mathbf{q}$ due to a virtual transition from a state $\mathbf{k}_{F}^{+}$
just above the Fermi energy. The total sum represents the reduction of the
occupation of the state $\mathbf{k}_{F}^{+}$ (before normalization).

\subsection{Satellite states in the ground state}

At $T=0$ in the absence of the electron-phonon interaction all states within
the Fermi sphere with $k\leq k_{F}$ ($k_{F}$ is the Fermi wave number) are
occupied and all other states are empty. We denote this ground state of the
electron system as $\left\vert \Psi_{0}\right\rangle =%
{\textstyle\prod\limits_{k^{\prime}<k_{F}}}
c_{\mathbf{k}^{\prime}}^{\ast}\left\vert \Phi_{0}\right\rangle $. We have
virtual electron-phonon emissions from the occupied state $\mathbf{k}^{\prime
}$ into the empty states $\mathbf{k}^{\prime\prime}$ emitting phonons $\left(
\mathbf{q},\lambda\right)  $. Now the state $\mathbf{k}^{\prime}$ makes a
transition into one (or several) states $\mathbf{k}^{\prime\prime}$ and
creates a phonon $\left(  \mathbf{q},\lambda\right)  $. The resulting state
can be described as
\begin{align*}
\widetilde{c_{\mathbf{k}^{\prime}}^{\ast}}  &  =\left(  c_{\mathbf{k}^{\prime
}}^{\ast}+%
{\textstyle\sum_{\mathbf{k}^{\prime\prime},\mathbf{q,}\lambda}}
\alpha_{\mathbf{k}^{\prime\prime},\mathbf{k}^{\prime}\mathbf{,q,}\lambda
}c_{\mathbf{k}^{\prime\prime}}^{\ast}a_{\mathbf{q},\lambda}^{\ast}\right) \\
&  =\left(  1+%
{\textstyle\sum_{\mathbf{k}^{\prime\prime},\mathbf{q,}\lambda}}
\alpha_{\mathbf{k}^{\prime\prime},\mathbf{k}^{\prime}\mathbf{,q,}\lambda
}c_{\mathbf{k}^{\prime\prime}}^{\ast}c_{\mathbf{k}^{\prime}}a_{\mathbf{q}%
,\lambda}^{\ast}\right)  c_{\mathbf{k}^{\prime}}^{\ast}%
\end{align*}
The amplitude of the satellites we denote as $\alpha_{\mathbf{k}^{\prime
\prime},\mathbf{k}^{\prime}\mathbf{,q}}$. This state is not normalized.

For the new ground state we make the product ansatz%
\begin{equation}
\widetilde{\Psi_{0}}\left(  t\right)  =\left[
{\textstyle\prod\limits_{\mathbf{k}^{\prime}}}
\left(  1+%
{\textstyle\sum_{\mathbf{k}^{\prime\prime},\mathbf{q,}\lambda}}
\alpha_{\mathbf{k}^{\prime\prime},\mathbf{k}^{\prime}\mathbf{,q,}\lambda
}c_{\mathbf{k}^{\prime\prime}}^{\ast}c_{\mathbf{k}^{\prime}}a_{\mathbf{q}%
,\lambda}^{\ast}\right)  \left\vert \Psi_{0}\right\rangle \right]
e^{-\frac{i}{\hbar}E_{0}t}\label{Psi0}%
\end{equation}
where $E_{0}$ is the new ground state energy. All products over $\mathbf{k}%
^{\prime}$ and summations $\mathbf{k}^{\prime\prime}$ are restricted to
$k^{\prime}<k_{F}$, $k^{\prime\prime}>k_{F}$. The hamiltonian is
\[
H=%
{\textstyle\sum_{\mathbf{p}}}
\varepsilon_{p}c_{\mathbf{p}}^{\ast}c_{\mathbf{p}}+%
{\textstyle\sum_{\mathbf{p}_{1}\mathbf{,p}_{2},\mathbf{q}^{\prime
}\mathbf{,\lambda}}}
g_{\mathbf{p}_{2}\mathbf{-p}_{1},\mathbf{q}^{\prime}\mathbf{,\lambda}%
}c_{\mathbf{p}_{2}}^{\ast}c_{\mathbf{p}_{1}}\left(  a_{\mathbf{q}^{\prime
}\mathbf{,\lambda}}+a_{-\mathbf{q}^{\prime}\mathbf{,\lambda}}^{\ast}\right)
\]
The Schroedinger equation is
\[
H\widetilde{\Psi_{0}}=E_{0}\widetilde{\Psi_{0}}%
\]
In order for $\widetilde{\Psi_{0}}$ in equ. $\left(  \ref{Psi0}\right)  $ to
be an approximate eigenstate of the hamitonian to first order in the
electron-phonon interaction $g_{\mathbf{p}_{2}\mathbf{-p}_{1},\mathbf{q}%
^{\prime}\mathbf{,\lambda}}$ the states $\left(  c_{\mathbf{k}^{\prime}}%
^{\ast}+%
{\textstyle\sum_{\mathbf{k}^{\prime\prime},\mathbf{q,}\lambda}}
\alpha_{\mathbf{k}^{\prime\prime},\mathbf{k}^{\prime}\mathbf{,q,}\lambda
}c_{\mathbf{k}^{\prime\prime}}^{\ast}a_{\mathbf{q},\lambda}^{\ast}\right)
\left\vert \Phi_{0}\right\rangle $ must be (approximate) eigenstates of the
hamiltonian, i.e.
\[
H\left(  c_{\mathbf{k}^{\prime}}^{\ast}+%
{\textstyle\sum_{\mathbf{k}^{\prime\prime},\mathbf{q,}\lambda}}
\alpha_{\mathbf{k}^{\prime\prime},\mathbf{k}^{\prime}\mathbf{,q,}\lambda
}c_{\mathbf{k}^{\prime\prime}}^{\ast}a_{\mathbf{q},\lambda}^{\ast}\right)
\left\vert \Phi_{0}\right\rangle =\eta_{\mathbf{k}^{\prime}}^{0}\left(
c_{\mathbf{k}^{\prime}}^{\ast}+%
{\textstyle\sum_{\mathbf{k}^{\prime\prime},\mathbf{q,}\lambda}}
\alpha_{\mathbf{k}^{\prime\prime},\mathbf{k}^{\prime}\mathbf{,q,}\lambda
}c_{\mathbf{k}^{\prime\prime}}^{\ast}a_{\mathbf{q},\lambda}^{\ast}\right)
\left\vert \Phi_{0}\right\rangle
\]
This yields
\begin{align*}
&  \eta_{\mathbf{k}^{\prime}}^{0}\left(  c_{\mathbf{k}^{\prime}}^{\ast}+%
{\textstyle\sum_{\mathbf{k}^{\prime\prime},\mathbf{q,}\lambda}}
\alpha_{\mathbf{k}^{\prime\prime},\mathbf{k}^{\prime}\mathbf{,q,}\lambda
}c_{\mathbf{k}^{\prime\prime}}^{\ast}a_{\mathbf{q},\lambda}^{\ast}\right)
\left\vert \Phi_{0}\right\rangle \\
&  =%
{\textstyle\sum_{\mathbf{k}^{\prime}}}
\left[
\begin{array}
[c]{c}%
\varepsilon_{\mathbf{k}^{\prime}}+%
{\textstyle\sum_{\mathbf{k}^{\prime\prime},\mathbf{q,}\lambda}}
\left(  \varepsilon_{\mathbf{k}^{\prime\prime}}+\hbar\omega\right)
\alpha_{\mathbf{k}^{\prime\prime},\mathbf{k}^{\prime}\mathbf{,q,}\lambda
}c_{\mathbf{k}^{\prime\prime}}^{\ast}c_{\mathbf{k}^{\prime}}a_{\mathbf{q}%
,\lambda}^{\ast}\\
+%
{\textstyle\sum_{\mathbf{k}^{\prime\prime},\mathbf{q,}\lambda}}
g_{\mathbf{k}^{\prime\prime}-\mathbf{k}^{\prime},-\mathbf{q,}\lambda
}c_{\mathbf{k}^{\prime\prime}}^{\ast}c_{\mathbf{k}^{\prime}}a_{\mathbf{q}%
,\lambda}^{\ast}\\
+%
{\textstyle\sum_{\mathbf{k}^{\prime\prime},\mathbf{q,}\lambda}}
g_{\mathbf{k}^{\prime}\mathbf{-k}^{\prime\prime},\mathbf{q,}\lambda
}c_{\mathbf{k}^{\prime}}^{\ast}c_{\mathbf{k}^{\prime\prime}}a_{\mathbf{q,}%
\lambda}\alpha_{\mathbf{k}^{\prime\prime},\mathbf{k}^{\prime}\mathbf{,q,}%
\lambda}c_{\mathbf{k}^{\prime\prime}}^{\ast}c_{\mathbf{k}^{\prime}%
}a_{\mathbf{q},\lambda}^{\ast}%
\end{array}
\right]  \\
&  \ast\left(  c_{\mathbf{k}^{\prime}}^{\ast}+%
{\textstyle\sum_{\mathbf{k}^{\prime\prime},\mathbf{q,}\lambda}}
\alpha_{\mathbf{k}^{\prime\prime},\mathbf{k}^{\prime}\mathbf{,q,}\lambda
}c_{\mathbf{k}^{\prime\prime}}^{\ast}a_{\mathbf{q},\lambda}^{\ast}\right)
\left\vert \Phi_{0}\right\rangle
\end{align*}
This yields%
\begin{equation}
\eta_{\mathbf{k}^{\prime}}^{0}=\varepsilon_{\mathbf{k}^{\prime}}+%
{\textstyle\sum_{\mathbf{k}^{\prime\prime},\mathbf{q,}\lambda}}
g_{\mathbf{k}^{\prime}\mathbf{-k}^{\prime\prime},\mathbf{q,}\lambda}%
\alpha_{\mathbf{k}^{\prime\prime},\mathbf{k}^{\prime}\mathbf{,q,}\lambda
}\label{E0}%
\end{equation}%
\[
\eta_{\mathbf{k}^{\prime}}^{0}\alpha_{\mathbf{k}^{\prime\prime},\mathbf{k}%
^{\prime}\mathbf{,q,}\lambda}=\left(  \varepsilon_{\mathbf{k}^{\prime\prime}%
}+\hbar\omega\right)  \alpha_{\mathbf{k}^{\prime\prime},\mathbf{k}^{\prime
}\mathbf{,q,}\lambda}+g_{\mathbf{k}^{\prime\prime}-\mathbf{k}^{\prime
},-\mathbf{q,}\lambda}%
\]
It follows that%
\begin{equation}
\alpha_{\mathbf{k}^{\prime\prime},\mathbf{k}^{\prime}\mathbf{,q,}\lambda
}=\frac{g_{\mathbf{k}^{\prime\prime}-\mathbf{k}^{\prime},-\mathbf{q,}\lambda}%
}{\left(  \eta_{\mathbf{k}^{\prime}}^{0}-\varepsilon_{\mathbf{k}^{\prime
\prime}}-\hbar\omega\right)  }\label{alf}%
\end{equation}
with the self-consistency condition%
\begin{equation}
\eta_{\mathbf{k}^{\prime}}^{0}=\varepsilon_{\mathbf{k}^{\prime}}+%
{\textstyle\sum_{\mathbf{k}^{\prime\prime},\mathbf{q,}\lambda}}
\frac{\left\vert g_{\mathbf{k}^{\prime}\mathbf{-k}^{\prime\prime}%
,\mathbf{q,}\lambda}\right\vert ^{2}}{\left(  \eta_{\mathbf{k}^{\prime}}%
^{0}-\varepsilon_{\mathbf{k}^{\prime\prime}}-\hbar\omega\right)  }\label{eth}%
\end{equation}

There are two approximations involved in the product ansatz: (a) The Pauli
principle excludes for $\widetilde{c_{\mathbf{k}_{1}^{\prime}}^{\ast}%
}\widetilde{c_{\mathbf{k}_{2}^{\prime}}^{\ast}}=$ $\left(  c_{\mathbf{k}%
_{1}^{\prime}}^{\ast}+%
{\textstyle\sum_{\mathbf{k}_{1}^{\prime\prime},\mathbf{q}_{1}\mathbf{,}%
\lambda}}
\alpha_{\mathbf{k}_{1}^{\prime\prime},\mathbf{k}_{1}^{\prime}\mathbf{,q,}%
\lambda}c_{\mathbf{k}_{1}^{\prime\prime}}^{\ast}a_{\mathbf{q}_{1},\lambda
}^{\ast}\right)  \ast$ $\left(  c_{\mathbf{k}_{2}^{\prime}}^{\ast}+%
{\textstyle\sum_{\mathbf{k}_{2}^{\prime\prime},\mathbf{q}_{2}\mathbf{,}%
\lambda}}
\alpha_{\mathbf{k}_{2}^{\prime\prime},\mathbf{k}_{2}^{\prime}\mathbf{,q,}%
\lambda}c_{\mathbf{k}_{2}^{\prime\prime}}^{\ast}a_{\mathbf{q}_{2},\lambda
}^{\ast}\right)  $ the double occupancy of the state $c_{\mathbf{k}%
_{1}^{\prime\prime}}^{\ast}c_{\mathbf{k}_{2}^{\prime\prime}}^{\ast}$ for
$\mathbf{k}_{1}=\mathbf{k}_{2}$. (b) The electron-phonon interaction can
introduce a transition from the state $c_{\mathbf{k}_{2}}^{\ast}$ into the
satellite of $\widetilde{c_{\mathbf{k}_{1}^{\prime}}^{\ast}}$ yielding a state
$c_{\mathbf{k}_{1}^{\prime\prime}}^{\ast}a_{\mathbf{q}_{1},\lambda}^{\ast
}c_{\mathbf{k}_{1}^{\prime}}^{\ast}a_{\mathbf{q}_{2},\lambda}^{\ast}$. This
state is neglected.

The total ground-state energy in this approximation is%
\[
E_{0}=%
{\textstyle\sum_{\mathbf{k}^{\prime}}}
\eta_{\mathbf{k}^{\prime}}^{0}%
\]
The occupation of the states $c_{\mathbf{k}^{\prime\prime}}^{\ast
}a_{\mathbf{q},\lambda}^{\ast}$ is given by%
\[
\left\vert \alpha_{\mathbf{k}^{\prime\prime},\mathbf{k}^{\prime}%
\mathbf{,q,}\lambda}\right\vert ^{2}=\frac{\left\vert g_{\mathbf{k}%
^{\prime\prime}-\mathbf{k}^{\prime},-\mathbf{q,}\lambda}\right\vert ^{2}%
}{\left(  \eta_{\mathbf{k}^{\prime}}^{0}-\varepsilon_{\mathbf{k}^{\prime
\prime}}-\hbar\omega\right)  ^{2}}%
\]
To normalize each state $\widetilde{c_{\mathbf{k}^{\prime}}^{\ast}}$ it has to
be divided by $\left[  1+%
{\textstyle\sum_{\mathbf{k}^{\prime\prime},\mathbf{q},\lambda}}
\frac{\left\vert g_{\mathbf{k}^{\prime\prime}-\mathbf{k}^{\prime}%
,-\mathbf{q,}\lambda}\right\vert ^{2}}{\left(  \eta_{\mathbf{k}^{\prime}}%
^{0}-\varepsilon_{\mathbf{k}^{\prime\prime}}-\hbar\omega\right)  ^{2}}\right]
^{1/2}.$

It should be emphasized that the energy $\eta_{\mathbf{k}^{\prime}}^{0}$ is
not the energy of a hole at $\mathbf{k}^{\prime}$. It is here only a
mathematical abbreviation.

\subsection{Ground state plus one electron}

Now we perform self-sonsistent perturbation calculation starting with the
unperturbed ground state plus one electron $\mathbf{k}_{0}$. We call this
state $\Psi_{0;\mathbf{k}_{0}}$ (with \ $\left\vert \mathbf{k}^{\prime
}\right\vert <k_{F}$)
\[
\Psi_{0;\mathbf{k}_{0}}=c_{\mathbf{k}_{0}}^{\ast}%
{\textstyle\prod\limits_{\mathbf{k}^{\prime}}}
c_{\mathbf{k}^{\prime}}^{\ast}\left\vert \Phi_{0}\right\rangle
\]
The resulting state in the presence of electron-phonon interaction is
\[
\widetilde{\Psi_{0;\mathbf{k}_{0}}}=\widetilde{c_{\mathbf{k}_{0}}^{\ast}}%
{\textstyle\prod\limits_{\mathbf{k}^{\prime}}}
\widetilde{c_{\mathbf{k}^{\prime}}^{\ast}}\left\vert \Phi_{0}\right\rangle
\]
with $\widetilde{c_{\mathbf{k}^{\prime}}^{\ast}}=\left(  c_{\mathbf{k}%
^{\prime}}^{\ast}+%
{\textstyle\sum_{\mathbf{k}^{\prime\prime},\mathbf{q,}\lambda}}
\alpha_{\mathbf{k}^{\prime\prime},\mathbf{k}^{\prime}\mathbf{,q,}\lambda
}c_{\mathbf{k}^{\prime\prime}}^{\ast}a_{\mathbf{q},\lambda}^{\ast}\right)  $
where $\mathbf{k}^{\prime\prime}\neq\mathbf{k}_{0}$.

When we derive the corresponding Schroedinger equations we obtain for
$\left\vert \mathbf{k}^{\prime}\right\vert <k_{F}$
\[
\alpha_{\mathbf{k}^{\prime\prime},\mathbf{k}^{\prime}\mathbf{,q,}\lambda
}=\frac{g_{\mathbf{k}^{\prime\prime}-\mathbf{k}^{\prime},-\mathbf{q,}\lambda}%
}{\left(  \eta_{\mathbf{k}^{\prime}}-\varepsilon_{\mathbf{k}^{\prime\prime}%
}-\hbar\omega\right)  }%
\]%
\[
\eta_{\mathbf{k}^{\prime}}=\varepsilon_{\mathbf{k}^{\prime}}+%
{\textstyle\sum_{\mathbf{k}^{\prime\prime},\mathbf{q,}\lambda}}
g_{\mathbf{k}^{\prime}\mathbf{-k}^{\prime\prime},\mathbf{q,}\lambda}%
\alpha_{\mathbf{k}^{\prime\prime},\mathbf{k}^{\prime}\mathbf{,q,}\lambda
}-g_{\mathbf{k}^{\prime}\mathbf{-k}_{0},\mathbf{q,}\lambda}\alpha
_{\mathbf{k}_{0},\mathbf{k}^{\prime}\mathbf{,q,}\lambda}%
\]
The amplitude $\alpha_{\mathbf{k}^{\prime\prime},\mathbf{k}^{\prime
}\mathbf{,q,}\lambda}$ has a slightly different energy denominator compared
with the ground state since $\eta_{\mathbf{k}^{\prime}}^{0}$ \ is replaced by
$\eta_{\mathbf{k}^{\prime}}$. However, the difference is of third order in the
electron-phonon matrix element and will be neglected in our approximation. For
the state $\widetilde{c_{\mathbf{k}_{0}}^{\ast}}=\left(  c_{\mathbf{k}_{0}%
}^{\ast}+%
{\textstyle\sum_{\mathbf{k}^{\prime\prime},\mathbf{q,}\lambda}}
\alpha_{\mathbf{k}^{\prime\prime},\mathbf{k}_{0}\mathbf{,q,}\lambda
}c_{\mathbf{k}^{\prime\prime}}^{\ast}a_{\mathbf{q},\lambda}^{\ast}\right)  $
one obtains in analogy%
\[
\alpha_{\mathbf{k}^{\prime\prime},\mathbf{k}_{0}\mathbf{,q,}\lambda}%
=\frac{g_{\mathbf{k}^{\prime\prime}-\mathbf{k}_{0},-\mathbf{q,}\lambda}%
}{\left(  \eta_{\mathbf{k}_{0}}-\varepsilon_{\mathbf{k}^{\prime\prime}}%
-\hbar\omega\right)  }%
\]%
\[
\]

The total energy of $\widetilde{\Psi_{0;\mathbf{k}_{0}}}$ becomes%
\begin{align*}
E_{0;\mathbf{k}_{0}}  &  =%
{\textstyle\sum_{\mathbf{k}^{\prime}}}
\eta_{\mathbf{k}^{\prime}}+\eta_{\mathbf{k}_{0}}\\
&  =E_{0}+\varepsilon_{\mathbf{k}_{0}}+%
{\textstyle\sum_{\mathbf{k}^{\prime\prime},\mathbf{q,}\lambda}}
g_{\mathbf{k}_{0}\mathbf{-k}^{\prime\prime},\mathbf{q,}\lambda}\alpha
_{\mathbf{k}^{\prime\prime},\mathbf{k}_{0}\mathbf{,q,}\lambda}-%
{\textstyle\sum_{\mathbf{k}^{\prime}}}
g_{\mathbf{k}^{\prime}\mathbf{-k}_{0},\mathbf{q,}\lambda}\alpha_{\mathbf{k}%
_{0},\mathbf{k}^{\prime}\mathbf{,q,}\lambda}%
\end{align*}
The quasi-particle energy $E_{\mathbf{k}_{0}}$ of the state $\mathbf{c}%
_{\mathbf{k}_{0}}^{\ast}$ is then%
\[
E_{\mathbf{k}_{0}}=\varepsilon_{\mathbf{k}_{0}}+%
{\textstyle\sum_{\mathbf{k}^{\prime\prime},\mathbf{q,}\lambda}}
g_{\mathbf{k}_{0}\mathbf{-k}^{\prime\prime},\mathbf{q,}\lambda}\alpha
_{\mathbf{k}^{\prime\prime},\mathbf{k}_{0}\mathbf{,q,}\lambda}-%
{\textstyle\sum_{\mathbf{k}^{\prime}}}
g_{\mathbf{k}^{\prime}\mathbf{-k}_{0},\mathbf{q,}\lambda}\alpha_{\mathbf{k}%
_{0},\mathbf{k}^{\prime}\mathbf{,q,}\lambda}%
\]
which yields%
\begin{equation}
E_{\mathbf{k}_{0}}=\varepsilon_{\mathbf{k}_{0}}+%
{\textstyle\sum_{\mathbf{k}^{\prime\prime},\mathbf{q,}\lambda}}
\frac{\left\vert g_{\mathbf{k}^{\prime\prime}-\mathbf{k}_{0},-\mathbf{q,}%
\lambda}\right\vert ^{2}}{\left(  \eta_{\mathbf{k}_{0}}-\varepsilon
_{\mathbf{k}^{\prime\prime}}-\hbar\omega\right)  }-%
{\textstyle\sum_{\mathbf{k}^{\prime}}}
\frac{\left\vert g_{\mathbf{k}_{0}-\mathbf{k}^{\prime},-\mathbf{q,}\lambda
}\right\vert ^{2}}{\left(  \eta_{\mathbf{k}^{\prime}}-\varepsilon
_{\mathbf{k}_{0}}-\hbar\omega\right)  } \label{Ek0}%
\end{equation}

If one compares this expression for $E_{\mathbf{k}_{0}}$ with the Green
function expression in equ. (\ref{Ek0GF}) one recognizes differences in the
energy denominators. In the Green function expression the $\mathbf{k}_{0}%
$-energy in the denominators is given by $E_{\mathbf{k}_{0}}.$ There the
transition from $\mathbf{k}^{\prime}$ to $\mathbf{k}_{0}$ is replaced by a
transition from $\mathbf{k}_{0}$ to $\mathbf{k}^{\prime}$. The physics behind
this is not obvious. As a matter of fact Schrieffer uses in his
Superconductivity book both approaches in parallel in the discussion of the
electron-phonon self energy.

\end{document}